\documentclass[apj]{emulateapj}

\usepackage{wasysym}
\usepackage{rotating}
\usepackage[dvips]{color}
\usepackage{verbatim}
\usepackage{soul}

\slugcomment{Accepted for publication in ApJ Main Journal.}

\shorttitle{Resolving the rotation measure of the M87 jet on kilo-parsec scales}

\shortauthors{Algaba, Asada \& Nakamura}

\begin{document}

\title{Resolving the Rotation Measure of the M87 jet on kilo-parsec scales}

\author{J. C. Algaba$^{1,2}$, K. Asada$^2$ and M. Nakamura$^2$}
\affil{$^1$Korea Astronomy \& Space Science Institute, 776, Daedeokdae-ro, Yuseong-gu, Daejeon, Republic of Korea 305-348 \\$^2$Academia Sinica, Institute of Astronomy and Astrophysics, 11F of Astronomy-Mathematics Building, AS/NTU. No.1, Sec. 4, Roosevelt Rd, Taipei 10617, Taiwan, R.O.C }

\begin{abstract}

We investigate the distribution of Faraday rotation measure (RM) in the M87 jet at arc-second scales by using archival polarimetric VLA data at 8, 15, 22 and 43 GHz. We resolve the structure of the RM in several knots along the jet for the first time. We derive the power spectrum in the arcsecond scale jet and find indications that the RM cannot be associated with a turbulent magnetic field with 3D Kolmogorov spectrum. Our analysis indicates that the RM probed on jet scales has a significant contribution of a Faraday screen associated with the vicinity of the jet, in contrast with that on kiloparsec scales, typically assumed to be disconnected from the jet. Comparison with previous RM analyses suggests that the magnetic fields giving rise to the RMs observed in jet scales have different properties and are well less turbulent than these observed in the lobes.

\end{abstract}


\section{Introduction}

M87 hosts a supermassive black hole (SMBH) with $M_{\bullet}=(3.2 - 6.6) \times 10^{9} M_{\odot}$ \citep{Macchetto77,Gebhardt11,Walsh13}. It is one of the closest active galactic nuclei (AGNs) at a distance $D=16.7$ Mpc with an extended jet. The origin of the jet emission is considered  to be synchrotron radiation from radio through optical to X-ray \cite[see e.g.,][]{Biretta91,Perlman01,Harris03}. With an angular scale of 81 pc arcsec$^{-1}$ \citep{Blakeslee09} and viewing angle of $\theta = 10^{\circ} - 19^{\circ}$ \citep[see e.g., ][]{Biretta99,Wang09}\footnote{Through this paper we use $\theta\sim15\degr$ \citep{Wang09}.}, the M87 jet has been intensively studied as one of the best references in AGN jets.

High dynamic range and high resolution images were firstly obtained with VLA more than thirty years ago \citep{Owen80}. Since then, observations have revealed detailed morphology of the M87 jet in radio and optical bands \citep[see e.g.,][]{Owen89,Perlman99}. Strong polarization reaching almost the maximum in synchrotron radiation ($>60$\%) has been found on in the jet at both radio and optical bands\citep{Owen90, Perlman99}, suggesting magnetic fields are well ordered. The VLBI--scales core, however, is highly unpolarized ($<0.4$\%) in radio bands  \citep{HL06}. 
Polarization seems to behave differently in optical and radio bands: magnetic vector orientations are perpendicular to the direction of the jet in knots HST-1, D, E and F\footnote{Here we follow the traditional notation \citep[see e.g. ][]{Perlman01} to name the different features in the kiloparsec scale of the M87 jet.} in optical but remains parallel in radio wavelengths \citep{Perlman99}. 

A rotation of the polarization angle of $\sim75$\degr \, between optical and 6 cm over 20 arcseconds in M87 was found by \cite{Schmidt78} and confirmed by \cite{Dennison80}. Rotation measure (RM) observations of M87 were performed with the VLA at four independent frequencies in the 5 GHz band by \cite{Owen90}, who found high values (RM$\sim1000-2000$ rad~m$^{-2}$) in the lobes and low (RM$\sim200$ rad~m$^{-2}$) in the jet. Their interpretation was the existence of a Faraday screen consisting on a thick layer enveloping the radio lobes with the jet lying mostly in front of it. \cite{ZT03} found RM of the order of $\sim9000$ rad~m$^{-2}$ at milli-arc second scales with indications of a change of sign, interpreted as due to hot gas and narrow line region clouds.

The study of the RM power spectrum can also provide insights in order to clarify the structure of the magnetic field beneath and, indirectly, probe the origin of the RM. For example, a Kolmogorov power spectrum with power index $a=11/3$ can be expected if the turbulence is hydrodynamic \citep{Kolmogorov41} or consisting on a MHD cascade of a turbulent and isotropic magnetic field \citep{GoldreichSridhar97}. On the other hand, if the magnetic field is not isotropic but there is an underlying ordered component, then the mean field will suppress the energy cascade along the direction of the magnetic field, effectively modifying the resulting spectral slope towards lower values \citep[see e.g.][]{Moffatt67}.

Efforts in this direction have been lead by e.g. \cite{VogtEnslin03,Guidetti08}, suggesting that Kolmogorov spectra is possible on large (intra-cluster) scales. This approach has been used extensively on e.g. \mbox{Hydra A}, \mbox{Abell 400}, \mbox{Abell 2634} \citep{VogtEnslin03}, \mbox{Abell 2255} \citep{Govoni06} or \mbox{Abell 2382}  \citep{Guidetti08}. In many cases, however, the derived indices were found to be relatively flatter, with the power index in the range 2.1 to 3.2. \cite{Guidetti12} suggested that an interesting possibility is that the magnetic turbulence is closer to two-dimensional, in which case $a=8/3$. %(Minter & Spangler 1996).
The Kolmogorov spectrum found in M87, $a=3$ on scales up to 10 arc seconds, is not consistent with a Kolmogorov index \citep[][R. Laing, private communication]{GuidettiPhD}, although still large enough to be able to rule out a 2D Kolmogorov at these scales.

We report here our study of the RM in the M87 jet via image stacking of a series of multi--frequency polarimetry on VLA archives. Compared with previous analysis, we use various epochs to be able to combine various VLA configurations, sensitive to both more extended and compact regions of the jet, as well as to provide long lever arms for the RM $\lambda^2$-law fit. Our aim is to probe the jet regions, concentrating on the RM properties of the M87 jet, in contrast with previous studies that focused on the M87 lobes, and investigate the RM origin, both from the perspective of the location of the Faraday screen as well as a possible magnetic configuration (turbulent/non-turbulent) giving rise to it.

We describe the archival data reduction in section 2. We present observational results including rotation measure, polarization and power spectrum in section 3.  Interpretation and origin of the RM is discussed in section 4. A summary of our results in section 5. In the Appendix, we discuss several simulations that we have performed to understand the impact of the beam and blanking on various possible configurations of power spectrum maps.

\section{Archival data reduction}
\label{datareduction}
We analyzed four different epochs of VLA multi--frequency polarization observations of M87 obtained from the archive\footnote{\url{https://archive.nrao.edu}}. We summarize the observations in Table \ref{M87Observations}, where we include the frequencies used, the VLA configuration and the observations date. Data were reduced in AIPS\footnote{Astronomical Image Processing System, developed and maintained by the National Radio Astronomy Observatory} using standard methods and maps were created with the AIPS task IMAGR using robust weighting. 

\begin{table}
\begin{center}
\caption{Observing Sessions.\label{M87Observations}}
\begin{tabular}{cccc}
\tableline
\hline
Obs. Code & Freq. (GHz) & VLA Conf. & Date\\
(1)&(2)&(3)&(4)\\
\tableline
AH822B & 8, 15, 22 & A& 2003 Aug 24 \\
AH822C & 15, 22, 43 & B& 2003 Nov 16 \\
AH862B & 8, 15, 22 & A& 2004 Dec 31 \\
AH862C & 15, 22, 43 & B & 2005 May 3 \\
\tableline
\end{tabular}
\end{center}
\end{table}

We corrected for the intrinsic antenna polarization (D--terms) with the sources 0521+166 and 1224+035 as calibrators and for the polarization angle using 3C~286 as calibrator. We used the final calibrated visibilities to obtain maps of the Stokes Q and U distributions. Those were used to construct the polarized flux ($p=\sqrt{Q^2+U^2}$) and polarization angle [$\chi=(1/2)\arctan(U/Q)$] maps with the AIPS task COMB. In all cases we compared the polarization vectors with 3C~286.  We then convolved the images at 8, 15, 22 and 43 GHz with a beam of FWHM of 0.35\arcsec. Final stacked Q and U maps were used to construct the polarization and polarization noise maps which were hereafter used to obtain the RM map via a modified version of the AIPS task RM. 

Although the root-mean-square (rms) of an off-source region in the map is typically used for estimating the noise levels, it has been shown that this is generally an underestimation for polarization, especially as it does not take into account the dispersion arising from D-terms calibration  \citep{Hovatta12}. These can be estimated by
\begin{equation}
\sigma_{D-terms}\simeq\frac{\sigma_{\Delta}}{\sqrt{N_{ant}N_{IF}N_{scan}}}\sqrt{I^2+0.3 I_{peak}^2}
\end{equation}
where $\sigma_{\Delta}$ is given by the uncertainty in the D-terms, $N_{ant}$ is the number of antennas, $N_{IF}$ is the number of sub-bands for a given observing frequency and $N_{scan}$ is the number of scans with independent measurements of the parallactic angle. In our case, we have $N_{ant}\sim26$, $N_{scan}\sim9$ for each epoch and we estimate $\sigma_{\Delta}\sim0.008$. In practice, this implies that the noise levels associated with polarization is roughly twice the estimated rms, in agreement with \cite{Hovatta12} and \cite{Mahmud13}

We also note that, although $\chi$ values will have an error associated with the absolute polarization angle calibration, this will be affecting all points in the map systematically in the same direction and thus will not introduce spurious RM gradients \cite[See e.g.,][]{Mahmud09,Hovatta12,Mahmud13}.

The longer baselines of the VLA A configuration provide us with major resolution, whereas the B configuration allows us to capture more diffuse emission. Thus, combining both of them we are able to capture the features of the M87 in a more efficient way. Furthermore, given the $\lambda^2$ dependence on the RM, a longer frequency space can dramatically increase the accuracy of the RM. For example, by adding 8 GHz to the triplet 15-22-43 GHz we increase the $\lambda^2$ space by a factor of 4, which in turn leads to a RM 4 times more accurate. With this is our mind, we combined the multi epoch data obtained from the archive.

Stacking of the data was performed both in the image as well as in the visibility domain. We note that the improvement in the rms of the stacked images was lower than expected ($\sim$ a factor of 2). We examined this and concluded that this was due to i) limitations in the dynamic range and ii) variability of the M87 core and HST-1 feature along the various epochs. The latter effect was clearly significant, specially when stacking in the visibility domain, preventing us to obtain a better stacked total intensity map, both in terms of SNR and morphology of the brighter features upstream the jet. In order to check for reliability, we compared the Stokes I, Q and U maps from the individual epochs and the stacked maps. We concluded that, whereas knots D, E, F, A, B and C are reliable, variability of the regions upstream the jet, makes the total and polarized flux of these unreliable. Hence, we will no longer discuss the physical properties of the M87 core and HST-1 hereafter in this paper\footnote{We refer to the extensive work on the RM in HST-1 by \cite{Chen11} for completeness.}.

We obtained the RM map of the stacked images and compared their features with the individual epochs. Overall, the map is qualitatively very similar, indicating that variability is negligible. This is expected as any variation in the course of the various observed epochs would arise either due to unrealistic fast variations (of the order of hundreds of parsecs per year) in the vicinity of M87, or effects much closer to us, which we consider not to be the case (see discussion below). Quantitatively the accuracy of the RM is greatly improved due to the stacking by more than a factor of two. We note that the improvement of the RM by additional integration time on the stacked data is amply overwhelmed by the inclusion of additional frequencies.

\begin{figure}
\includegraphics[angle=0,scale=0.33,trim=0cm 0cm 0cm 0cm,clip=true]{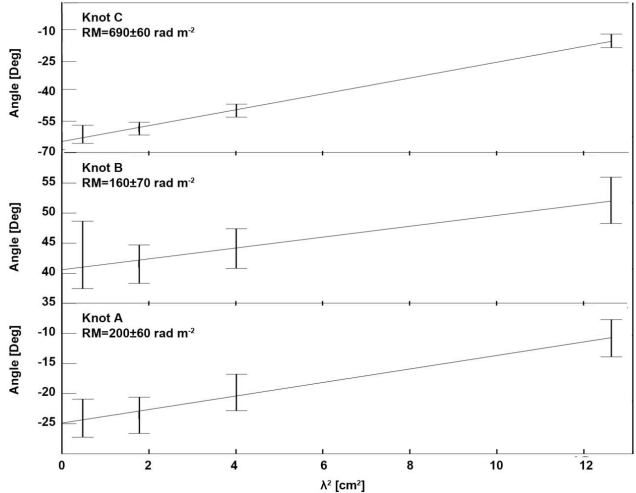}
\caption{Illustration of rotation measure fits for typical regions on knots A, B and C.} 
\label{RMfit}
\end{figure}

Uncertainties were calculated with the task RM based both on the polarization angle uncertainties (assuming a 2$\times$2 \degr error from polarization angle, plus correction due to Rician bias \citep{WardleKronberg74}), and the quality of the fit. We blanked these output pixels where the RM uncertainty exceeds $\sim100$ rad~m$^{-2}$, based both on estimations from the $\lambda^2$ range and maximum values which do not lead to spurious features along the jet. Some sample fits are shown in Figure \ref{RMfit}.

\section{Results}

\subsection{Rotation Measure and Polarization}

\begin{figure*}
\includegraphics[angle=0,scale=0.84,trim=0cm 0cm 0cm 0cm,clip=true]{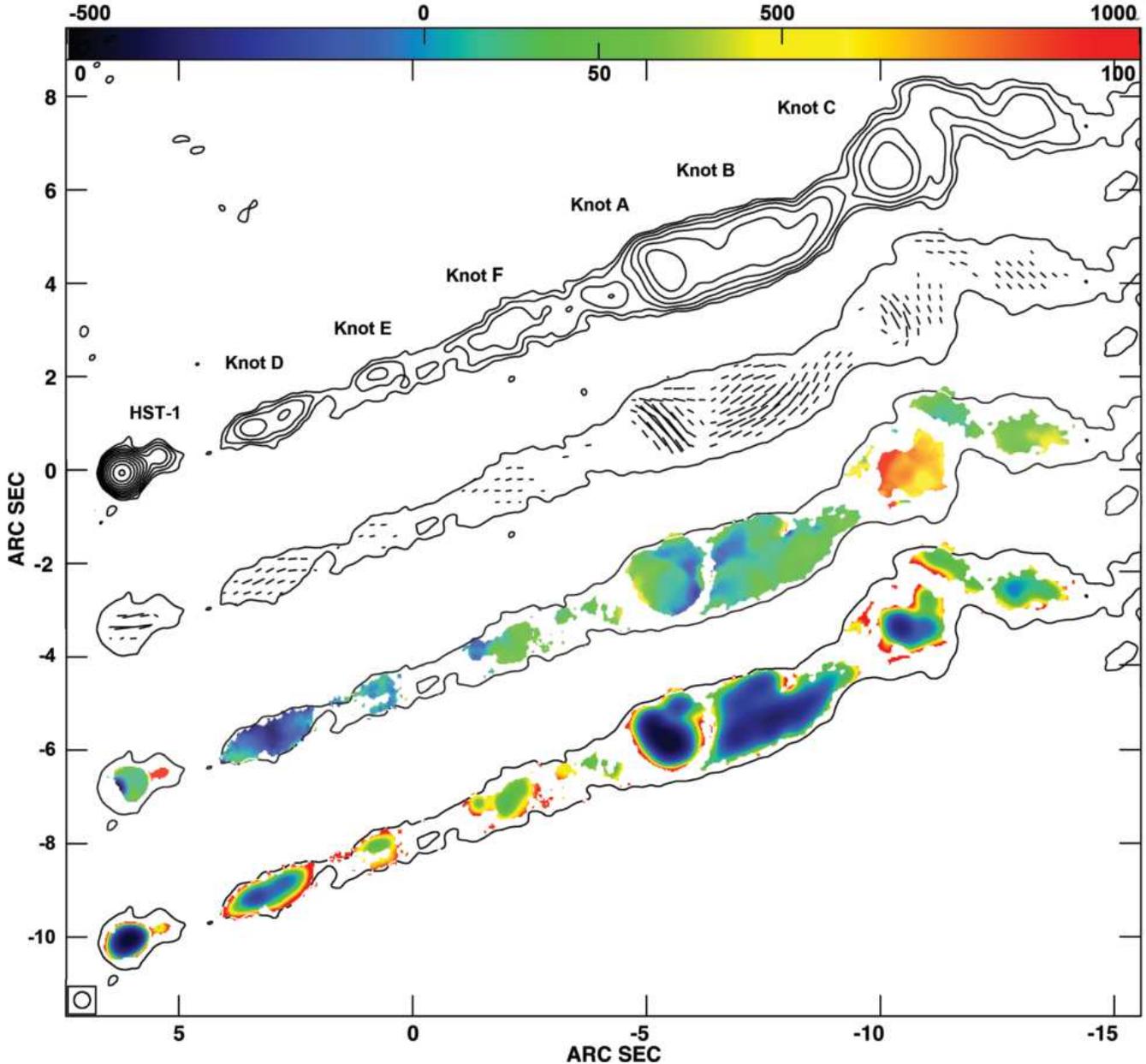}
\caption{Map of the kilo parsec scale M87 jet, with the main features (HST-1 and knots A, B, C, D, E) indicated. From top to bottom: i) total intensity 8 GHz map. Contours start at 3$\times$RMS with RMS=7.8$\times10^{-4}$ Jy/beam and increase in steps of (1, 2, 4, 8, 16, 32, 64, 128, 256, 512). ii) Magnetic vector polarization angle. Sticks indicate the Faraday corrected magnetic field direction with length proportional to the degree of polarization as 1 arc sec=40 mJy/beam. iii) Rotation measure map, with the color scale shown on top of the color bar, running from -500 to 1000 rad~m$^{-2}$. iv) Rotation measure errors map, with color scale shown on the bottom of the color bar, running from 0 to 100 rad~m$^{-2}$. }
\label{RMmap}
\end{figure*}

Resulting maps are presented in Figure \ref{RMmap}, where we show total intensity map for 8 GHz, Faraday corrected magnetic vectors (i.e, Faraday corrected electric vectors rotated 90\degr), and RM distributions and its error. Rotation measure is detected virtually all over the jet, specially in knots D, F, A, B and C. Except some small patches in knot E, F and outer regions of C, the regions where the RM is found extend at least one beam size. In particular this is remarkable for knots A, B and C, where RM is found to extend several beams both along and across the jet. 

Given the relatively higher frequencies used here compared with previous works, we can aim to the jet characteristics of the M87 RM, in contrast to other works that mainly focused on the larger scale lobes of M87. Furthermore, albeit the higher frequencies, the large range used allowed the RM errors to be tighter than in previous works. We are capable of resolving for the first time the M87 RM morphology on jet scales, and the robust detection of RM all along the jet, allows us to are able to utilize extended tools to provide detailed and comprehensive study of its properties.

RMs of several hundreds of rad~m$^{-2}$ are found in knots D, E, A and tail of knot C, in agreement with \cite{Owen90}. The RM appears to be larger on knot C, with values up to $\sim$1000 rad~m$^{-2}$. Inspection of the large scale RM map in \cite{Owen90} also shows a significant increase of the RM around knot C. As they are able to detect RM over the extended emission of the lobe, it appears that a structure giving rise to a larger RM is indeed crossing knot C. The only regions, other than knot C, where RM of the order of few thousands have been found are  the HST-1 complex \citep{Chen11} and upstream the jet, on VLBI scales \citep{ZT02}.

Although we are able to resolve the M87 RM, its structure seems to be quite smooth, and there is no clear overall morphology nor evident large fluctuations in the values of nearby (within a beam size) regions. Knot A seems to show some indications of a smaller RM downstream the jet, whereas knot D seems to have also lower RM in its centre. The RM values seem to be larger on the upper side of knot C than on the lower side. This is not clearly seen in any of the other knots in the M87 jet. Although knot A also shows some indications of a smaller RM on the lower side, this is not very clear and, considering various robustness criteria discussed by \cite{Hovatta12} or \cite{Algaba13}, we cannot find a robust indication of a RM gradient.

Faraday--corrected polarization (see Figure \ref{RMmap}) tends to be well aligned with the direction of the jet along most of its structure. Where the jet bends after knot B, polarization follows this bend. Two exceptions are knots A and C, where polarization appears almost perpendicular to the direction of the jet. This is in agreement with previous results by \cite{Owen89}.

\begin{figure}
\includegraphics[angle=0,scale=0.33,trim=0cm 0cm 0cm 0cm,clip=true]{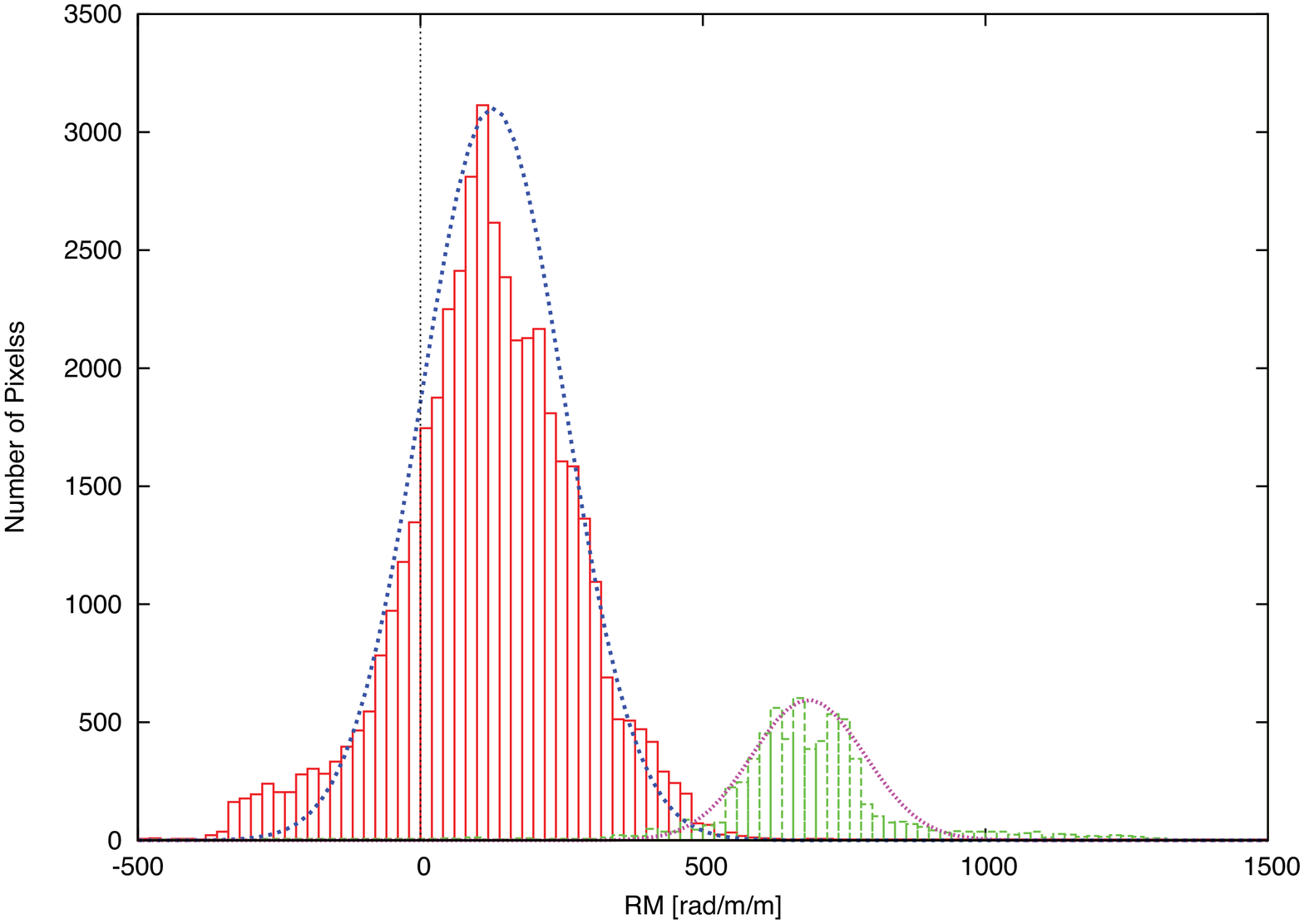}
\caption{Histogram of the RM values found in the M87 jet. Dashed green indicates values of the RM from knot C whereas solid red indicates values from the rest of the jet considered here. The vertical dotted line shows a null value for the RM. Two independent gaussian fittings are also shown.}
\label{RMhistogram}
\end{figure}

A histogram of the RM values is shown in Figure \ref{RMhistogram}. As discussed above, values from the core or HST-1 have not been included. 
There appear to be two clearly differentiated peaks in the RM distribution. Guided by the spatial distribution of RM values, we split the contribution from knot C and the rest of the jet. Indeed, it appears that most of the larger RM values are originated within knot C, whereas the rest of the jet contains solely lower values.
A gaussian distribution can be used to represent these if we consider to arise from two distributions. The RM in knot C would have a mean $<$RM$>_{knotC}$=680 rad~m$^{-2}$ with a standard deviation $\sigma^{RM}_{knotC}$=180 rad~m$^{-2}$, whereas for the rest of the jet $<$RM$>_{rest}$=130 rad~m$^{-2}$ and $\sigma^{RM}_{rest}$=120 rad~m$^{-2}$.

\subsection{Power Spectrum}

We now consider the power spectrum of the RM. We first blanked the regions corresponding to the M87 core and HST-1 of the RM map and pixels with huge differences and scatter on their RM values were removed from the data to decrease the presence of artefacts. We then calculated the Fourier transform and took the absolute square value to get the 2--dimensional power spectrum. We finally averaged over rings in the Fourier space to obtain the 1--dimensional power spectrum. To check the consistency of our results, we performed several tests. First, we subtracted the mean value of the RM from the RM map. Second, we obtained the power spectrum of the RM with different beam and cell sizes. In all these cases, the overall shape of the power spectrum did not differ significantly, although we found some small variation ($<$8\%) on the quantitative results. 

\begin{figure}
\includegraphics[angle=0,scale=0.91,trim=0cm 0cm 0cm 0cm,clip=true]{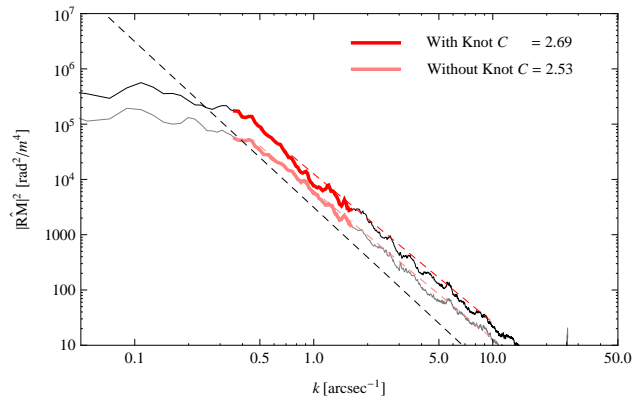}
\caption{Power spectrum of the RM for the M87 jet. The black curve, red bold curve and broken red line show the spectrum, reliable scales and power index fit for the RM in the M87 jet respectively when excluding knot C. The grey curve, pink bold curve and broken pink line are analogous when including knot C. The dashed black line shows a power index a=3 for eye guidance.}
\label{PowerSpectrum}
\end{figure}

The one--dimensional power spectrum (for both excluding and including the high RM values on knot C) is shown in Figure \ref{PowerSpectrum}.  First, let us discuss how to interpret this data. On one hand, it is clear that power spectrum values associated with regions in the RM map smaller than the beam size will not have a proper physical meaning. Strictly speaking, the minimum resolvable size will not be given only by the beam size, but will also be dependent on the SNR of the map \citep[see e. g., ][]{Lobanov05}. Given the complexity of a variable SNR over different regions of the map, we will follow here the more conservative approach and will consider that the power spectrum is not physical on Fourier scales larger than $\sim1/(2\times$beam~size$)\sim1.5$ arcsec$^{-1}$, given by the Nyquist sampling. On the other hand, due to the limited extension of RM, the largest separation between RM points gives us a lower limit of $\sim$0.3 arcsec$^{-1}$ in the power spectrum.

In order to estimate the power spectral index accuracy, we performed a Montecarlo simulation, where we took our initial RM map as our seed and pixel by pixel added a random gaussian value with zero mean and standard deviation given by the RM error at the given point. The final spectrum index error is given by the standard deviation of all the indices obtained for 100 realizations. We estimate it to be 0.12 for both cases (with or without accounting for knot C). We note that this is slightly smaller but comparable with the variations in the simulations discussed in the Appendix.

The reliable region of the power spectrum can be fitted with a power law spectrum $P\propto k^{-a}$ with index $a=2.5\pm0.1$, if we exclude the data of knot C; and $a=2.7\pm0.1$ otherwise\footnote{We do not fit a power law spectrum for knot C alone given that the number of data is too small and the fluctuations in its power spectrum too large to produce any reliable value.}. The difference in these values is compatible with the nominal error and thus we will not consider it significant. This result is much flatter than typical values of $a=11/3$ for a Kolmogorov--like spectrum. Previous works \citep[see e.g.,][]{VogtEnslin03,Laing08} have also found index significantly different from Kolmogorov in some other radio sources using different approach based on the real space study of the structure function.

In order to investigate to what extend the power law index $a\sim2.6$ could partially arise from possible effects or artefacts due to the limitation of the observations, we have simulated different RM map configurations (see  Appendix). Models include gaussian random fields with various initial power index to which we have applied different smoothing with a gaussian convolution and blanking. Our results indicate that, once the reliable regions of the power spectrum are taken into account, smoothing due to e.g., convolution of the beam, does not significantly affect the power index but blanking does flatten the spectrum, as the blanking can be modeled in the spectral domain as a window function causing some spectral leakage.

As discussed in the Appendix, this blanking--based flattening can be dramatic for a steep spectrum, but not very significant for a flatter one. In this sense, we estimate that with an observed power spectrum of $a\sim2.6$ the intrinsic power spectrum should not differ a great amount; whilst an observed spectrum of $a\gtrsim3.1$ ($>4\sigma$ larger than the actual observed one), should be necessary to consider the possibility of an intrinsic Kolmogorov spectrum. Thus, based on these simulations we conclude that the intrinsic M87 RM power spectrum may be slightly steeper than the observed one, but  incompatible with a 3D Kolmogorov phenomenology and different from the one found in the radio--lobes at larger scales.

\section{Interpretation}

\subsection{Origin of the Rotation Measure}

\subsubsection{Location of the Faraday Screen}
Polarization angles seem to follow the $\lambda^2$ linearity quite well (see Figure \ref{RMfit}). Also, the polarization angle rotates over more than $\sim$45\degr in knot C through the range of observed frequencies. These two properties seem to indicate that the origin of the RM is external (i.e, the rotating magnetized plasma is not intermixed with the emitting non-thermal electrons), although some contribution from internal Faraday rotation cannot be ruled out. Hence, we consider two possible origins for the Faraday screen giving rise to the observed RM: due to an unrelated external screen or related with the vicinity of the M87 jet.

With a galactic latitude $b=74.5\degr$ for M87, little contribution ($\lesssim10$s of rad m$^{-2}$) is expected from the interstellar medium in our galaxy \citep[see e.g, ][and references therein]{Pushkarev01,Simard-Normandin81}. The galactic environment of the Virgo cluster could also be a source for the rotation but, as already noted by \cite{Owen90}, M84, also located in this cluster, shows much smaller RM ($\sim30$ rad m$^{-2}$). Extreme values for Faraday rotation have been found not to exceed $\pm30$ rad m$^{-2}$ in a sample of large Virgo spirals \citep{Wezgowiec12}. Although these sources are in nature different to M87, this may indicate that the observed RM is likely linked to M87 itself.

A RM associated with a foreground external medium is one of the most common interpretations for integrated RM observed with VLA \citep[see e.g.,][]{Simard-Normandin81,Laing84,Andernach92,Pushkarev01}. \cite{ZT03} discussed about the possibility of the Faraday screen on much smaller (mas) scales to be the hot gas and narrow line region clouds. We consider that the accretion flow is unlikely to be the Faraday screen on the scales studied here. If this were the case, the dominant region for the flow should not extend to scales much larger than the Bondi radius which, for the case of M87, is located at $\sim200$ pc from the central engine, near the HST-1 structure, well within the regions upstream the jet we excluded from our study. Furthermore, a $|RM|\sim 10^{5-6}$ rad~m$^{-2}$ may be expected in the accretion flow in low-luminosity AGNs, such as M87 \citep{Kuo14}, compared with the derived value of $|RM|\sim 10^{2-3}$ rad~m$^{-2}$.

Strong winds from accretion flow can also be a possible source for the observed RM. Indeed, in some cases, winds in BAL QSOs have been observed reaching up to hundred of pc away from the nucleus \citep{Borguet13,Feruglio15}. For M87 we note however that, with X-ray temperatures $T\sim0.8 - 1$ keV at around the Bondi radius \citep{DiMatteo03} and stellar velocity dispersion of the bulge $\sigma\sim360$ km~s$^{-1}$ \citep{Gebhardt11}, a virial state $kT \sim \mu m \sigma^2$ is fairly holding, indicating that the ISM around the Bondi radius is gravitationally bounded and no winds beyond these scales are expected. Furthermore, if we assume $n_e\sim0.001 - 0.01$cm$^{-3}$ for a hypothetical wind (given that the wind density should be lower than that of the accretion flow) and an integration path of $L \sim 0.1$ kpc, we estimate a magnetic field of few mG for such wind, in order to obtain the observed RM values. This is similar or larger than the values of the magnetic field found in the M87 VLA jet itself \citep{Owen89}. Hence, we feel it is unlikely that a wind is the source of the observed RM.

\subsubsection{Nature of the RM}

The gaussian distribution of the RMs (see Figure \ref{RMhistogram}) seems to be in agreement with an scenario where the RM is generated by a turbulent and isotropic magnetic field along the line of sight. As the distribution is not centered at zero, this seems to indicate a larger scale magnetic field structure. If this is the case, the double peak of the RM distribution (or, alternatively, the existence of the larger RM region crossing knot C in \cite{Owen90}) could be explained by the existence of various of these larger scale fields. 
In contrast, it appears to be clear that the observed power spectrum is much flatter than the expected Kolmogorov spectrum, such as the one observed on cluster scales \citep[e.g. 3C~31, ][]{Laing08}, that would naturally arise under these circumstances\footnote{Note that Kolmogorov theory cannot be strictly applied to the case of intergalactic magnetic fields, as homogeneous and incompressible turbulences, which is assumed there, have been proven to be not always the case.}.

Another possibility, as previously speculated by \cite{Guidetti12}, is that the magnetic turbulence is two--dimensional in nature. The power index we obtain $a\sim2.6$ is indeed in agreement with the value $a=8/3$ expected for a 2D Kolmogorov turbulence. We suggest however that this is not the case here. As indicated by \cite{MinterSpangler96}, if we consider a thin 2D sheet to be the source of the jet RM, we would be probing the isotropic Kolmogorov turbulence for angular separations corresponding to distances \emph{smaller} than the thickness of the sheet, whereas we would expect the two--dimensional  turbulence for angular separations \emph{larger} than the thickness. If we consider the power index $a\sim3$ found by \cite{GuidettiPhD}, the case in M87 is totally reversed, as we see the power index to be smaller at \emph{smaller} scales.

It is quite reasonable that the observed RM may be the result as a combination of various independent magnetic distributions. One of the simplest cases to be considered would be the one where the observed RM distribution is the result of an initial Faraday rotation due to some ordered magnetic fields in the surroundings of the jet plus an additional rotation due to some random field in the lobes and/or the interstellar media. A very simple toy model to consider such case would be the combination of a Kolmogorov distribution plus an additional flatter contribution arising from a more ordered field. If we consider such case, the resulting power spectrum will be between that of these limiting cases, in the most general case larger than the observed one, depending on the different weighted contributions. 

We conjecture that the Faraday screen giving rise to the observed RM is not associated with a foreground external medium, and in turn we speculate that they are associated with the vicinity of M87 jet. If so, it is clear that its physical properties will not be easily described by a simple Kolmogorov model. The isotropic and incompressible conditions assumed in the Kolmogorov theory cannot be found in the jet environment, where knots, and MHD compressions are easily found. On the other hand, the presence of an ordered magnetic field in the jet would easily decrease the strength of the interactions flattening the spectral slope towards the values observed here.

Additionally, it is remarkable that the power index found on jet scales in this work $a\sim2.6$ is different from that on larger scales $a\sim3$. This seems to indicate that the regions giving rise to the observed RM can be different in nature and their magnetic field properties might also be unalike. The power index progressively deviating away from a Kolmogorov distribution as we probe smaller scales may be an indication of a gradually ordering of the magnetic field properties as we approach the comparatively more collimated jet, as expected from the high degree of polarization and well aligned polarization angles in the jet both in optical and radio wavelengths \citep[see e.g.][]{Perlman99}.

\subsubsection{Possible Scenario}

Following \cite{Owen90}, a possibility is that a fraction of the lobe is in the jet foreground towards our line of sight. In this scenario, the observed RM is partly generated by a screen associated with the jet, plus a non negligible contribution of RM from the lobes. If we follow this model, a superposition of a significant proportion ($\geq50$\%) of a Kolmogorov-like gaussian RM distribution to an otherwise uniform RM distribution will still exhibit a power index distinct of the later, within the errors. Taking this into account, and considering the relative values of the RM in the M87 jet and radio lobes, we estimate that a fraction of about $\lesssim10$\% of the lobe can be located in front of the jet with respect to our line of sight, with a contribution of possibly a similar order of magnitude as that of the jet.

In some regions of the jet, filaments or layers of the lobes with longer path length and/or locally over--dense electron density or magnetic field may be located between the jet and our line of sight. The RM in these regions will be probing a much larger contribution from the lobes and hence its properties, such as the RM value or power spectrum, will be closer to these in the radio lobes. This may be the case, for example, in knot C. Unfortunately, given the low polarization on the regions around knot C, as discussed above, it is difficult to make a connection between the RM values found in knot C and the rest of the jet. We remind that indications of such a filament crossing knot C are seen in \cite{Owen90}. 

At larger scales, in the lobes, the magnetic configuration will be such that the turbulence will be much stronger compared to the jet, and thus the observed spectrum will be much closer to the Kolmogorov type, as observed by \cite{GuidettiPhD}. On the other hand, \cite{Young02} found rich gas features, with the gas within the inner 3.5 kpc being at least a two--temperature plasma and \cite{Guidetti08} found evidence for radial scaling of gas, shells of compressed gas and shocks. It is thus clear that a simple Kolmogorov model will not fully describe the physical properties of the plasma producing the RM in the M87 radio lobes, and some other, more elaborated model, is necessary.

\subsection{No Evidence of a RM Gradient?}

Several mechanisms possibly associated with the physical properties in vicinity of the jet can produce a distinct morphology in the observed rotation measure in the form of a gradient. In the most simple cases, this will be induced because at least one of the magnetic field or the electron density inducing the Faraday rotation will suffer variations in its values. For example, it is expected that both magnetic fields and electron density will decrease along the jet \citep[e.g.][]{BK79}, causing in turn a decrease of the RM along downstream the jet. Alternatively, local enhancements of the magnetic field via a shock or a compression \citep{Laing80} can also produce a longitudinal RM gradient. Magnetic kinks, instabilities or jet bends may too produce a local RM increase.

If the jet contains a helical magnetic field, its toroidal component will produce a gradient, and possibly a sign reversal, on the RM across the jet,  \citep{Blandford93}. Such gradients have been observed in a variety of sources \citep[see e.g.,][among others]{Asada02,Gabuzda04,ZT05,Gomez08,Mahmud09,Croke10,Hovatta12,Algaba13}.  RMs are observed systematically on parsec scales of AGN \citep{Hovatta12} and have been detected on up to kilo parsec scales \citep{Kronberg11} although, due to polarization errors and beam effects, the limits of their reliability and interpretation are still under debate. 

Interestingly, no robust transverse RM gradient is observed in the M87 jet (nor has it been found in previous works by e.g. \cite{Owen90,ZT02,Hovatta12} at any scale). This is puzzling, as several studies seem to indicate that magnetic fields may play an important role in the M87 jet configuration \citep{Nakamura10,Hardee11,Nakamura14}, and several authors have suggested the existence of helical magnetic fields in its jet \citep[e.g.][]{Chen11,Nakamura14}, thus making M87 a priori a very good candidate to search for such gradients.

We note that so far we can not totally exclude the existence of a RM gradient transverse to the M87 jet with current observations due to the size of the errors or optimization of the observations for the RM properties on other non--jet regions. It may be possible that a moderate gradient that we are currently unable to identify may still exist. Such mild gradients can be obtained in some models for certain configurations of jet bulk velocities, viewing and magnetic fields pitch angles compatible with the M87 jet. \cite{BroderickLoeb09} found that even moderate relativistic motion produces dramatic alterations of the RM profile with respect to the static case, including in some cases a severe flattening or even reversal of the gradient. This is the case for the M87 jet, where knots studied here can reach apparent velocities between 0.4 and 1.6~c \citep{Meyer13}. Alternatively, if the observed RM is not entirely associated with the M87 jet, but contaminated by emission from the lobes, with a more tangled magnetic field configuration, any RM gradients intrinsically associated with the vicinity of the jet may be blurred of wiped out.

\section{Conclusions}

We have collected archival data to study multifrequency (8+15+22+43 GHz) VLA polarization images of the M87 jet. By image stacking, the use of a wide range in the $\lambda^2$ domain and various VLA configurations in order to trace both the extended and high resolution phenomenology of the jet, we are able to spatially resolve, for the first time, the rotation measure properties of the M87 jet on various knots at arcsecond scales.

Rotation measure is of the order of few hundreds of rad~m$^{-2}$, with some larger values up to $\sim$1000 rad~m$^{-2}$ in knot C, in agreement with previous results by \cite{Owen90}. The jet RM distribution can be fitted with two gaussians, one gaussian profile is caused almost entirely by the high values across knot C, while the other corresponds to the values in the rest of the jet. Interestingly, both gaussian profiles appear to have a similar standard deviation, of the order of $\sigma^{RM}\sim$120--180 rad~m$^{-2}$.

The power spectrum of the RM seems to follow a simple power law with index $a\sim2.6$ where reliable. This value is much lower than the one expected by a simple Kolmogorov distribution, based on an intergalactic turbulent magnetic field. Our results on the M87 RM distribution are not, in fact, consistent with a Kolmogorov index. Furthermore, the RM power spectrum index is different from the one found in the lobes at larger scales. These results seem to indicate that with our RM observations we are  probing different regions, possibly closer to the jet, with different magnetic properties.

We discuss the possible location and properties of the Faraday screen giving rise to the observed RM. Based on the RM distribution and the power spectrum, we suggest that the screen is not associated with the inter--cluster medium but much closer to the vicinity of the jet. A possible scenario to interpret our results is one where the RM screen is mostly associated with the sheath of the jet, (but not necessary linked with the spine helical magnetic field), while contributions arising from the lobes may not be negligible.

We suggest that the M87 jet contains a well ordered magnetic field, which dissipates into a more turbulent field towards the large scale lobes. The path length of the lobes is much larger than the one of the jet, giving rise to RM values about an order of magnitude larger. A small fraction of the lobes ($\lesssim10$\%) is located in front of the jet, and thus the observed RM results as a combination of contributions from both the jet and partial contamination from the lobes. A filament or structure arising in the lobes and crossing knot C along our line of sight could explain its apparently different RM properties.
\\[10pt]

\acknowledgments

We are grateful to D. Gabuzda, J. L. G\'omez, R. Laing, M. Inoue, P. T. P. Ho and the other members of the ASIAA VLBI group for valuable discussion and the anonymous referee for useful comments. The National Radio Astronomy Observatory is operated by Associated Universities, Inc., under contract with the National Science Foundation.

%{\it Facilities:} \facility{VLA}.

\appendix

We discuss here the dependence of the power spectrum with the RM map characteristics in order to understand its reliability. Our initial hypothesis is that  the parent distribution of the observed RM map is Kolmogorov in nature. This can be the case if the magnetic structure giving rise to such RM is a turbulent stochastic field. If so, the resulting distribution will be that of a gaussian random field. The power spectrum of the derived RM will be that of a Kolmogorov distribution with index $a=11/3$.

We performed a series of simulations to generate such gaussian random field maps. For our initial models, we initially produced maps in a $2048\times2048$ grid (equal to the map size in our observation) and derived power spectra with the following indices: $a=1.6$, $a=2.5$ and $a=3.6$. We find that a variance of about $\sim2$\% on the spectral power arises as a result of different realizations of our simulations. We then conducted the following simple tests: we i) convolved these maps with a gaussian to simulate the effects of the interferometric beam and ii) blanked the resulting maps. These steps were repeated various times in order to check for the reliability of the procedure and stability of the results.

\begin{figure*}[h]
\begin{center}
\includegraphics[angle=0,scale=0.5,trim=0cm -0.5cm 0cm 0cm,clip=true]{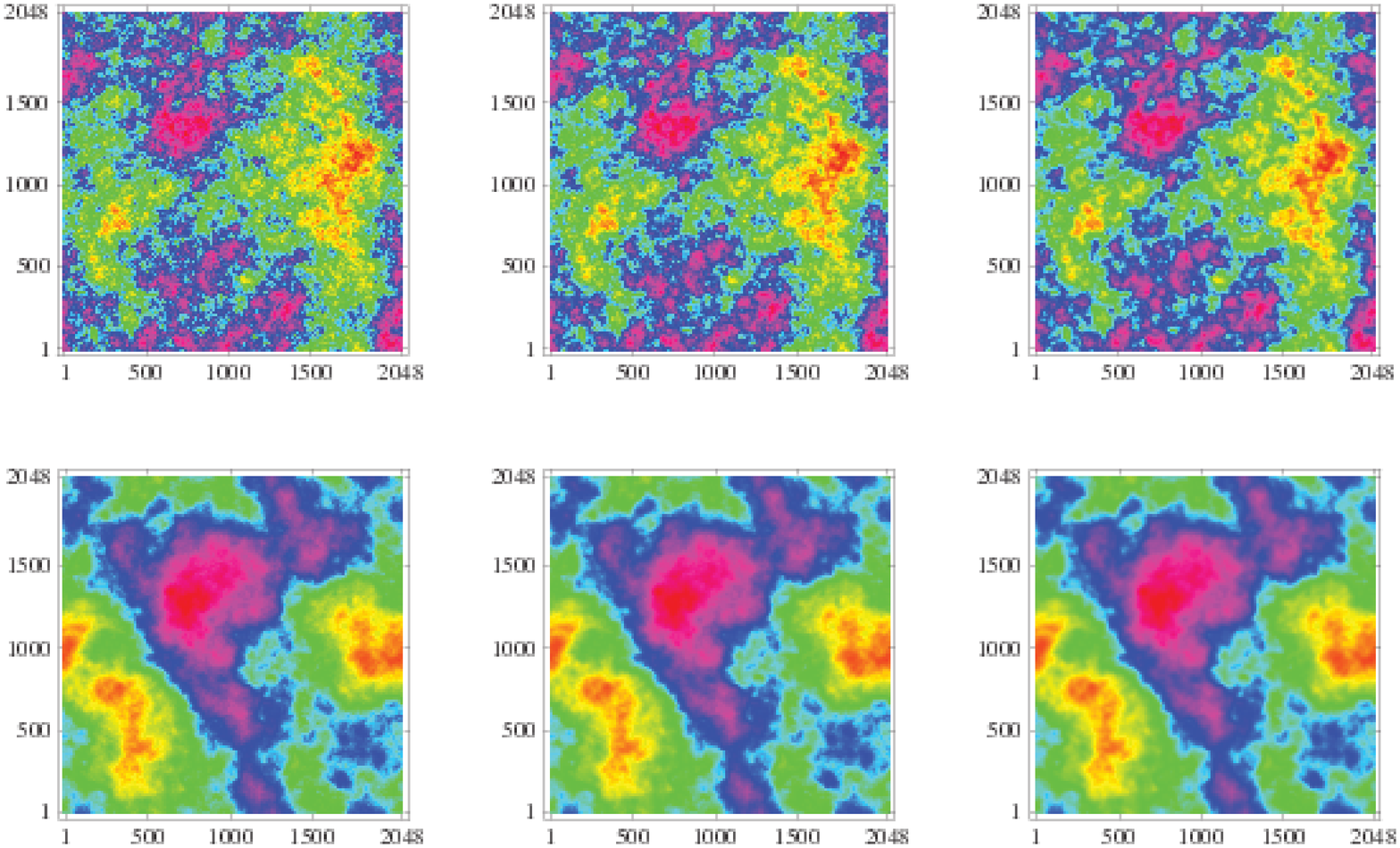}
\includegraphics[angle=0,scale=1,trim=0cm -0.5cm 0cm 0cm,clip=true]{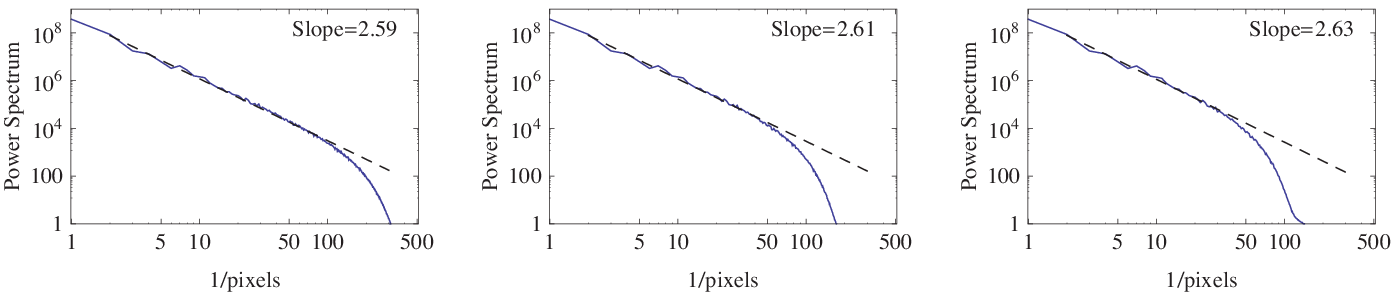}
\includegraphics[angle=0,scale=1,trim=0cm 0cm 0cm 0cm,clip=true]{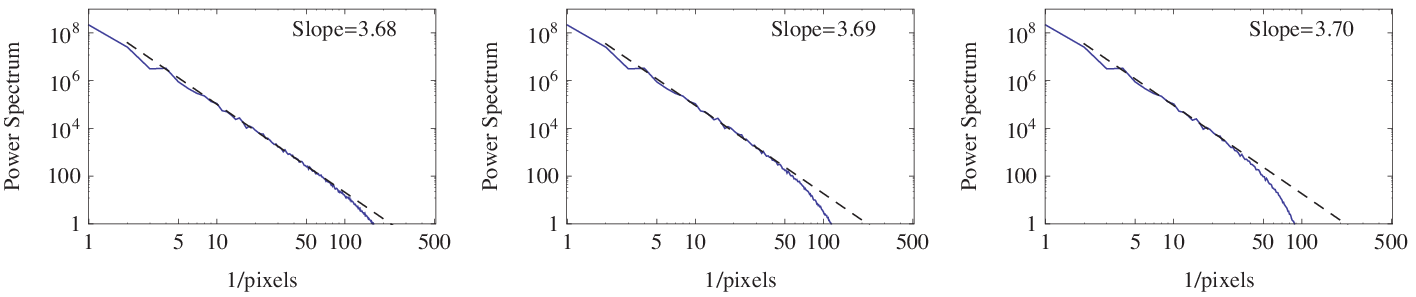}
\caption{Power spectrum of a single realization of random gaussian field with initial power index $a=2.5$ (top) and $a=3.6$ (bottom) convolved with a gaussian of 5 (left), 10 (middle) and 15 (right) pixels respectively. Resulting power index fitted to the reliable region of the slope is shown in the top right corner of each graphic. Fit is shown by the dashed line.}
\label{simulations1}
\end{center}
\end{figure*}

We convolved the maps with a gaussian with FWHM of 5, 10 and 15 pixels, which correspond in the Fourier space to $\sim$ 205, 100 and 68 pixels$^{-1}$, respectively. Power spectra of a subset of the synthesized maps are shown in Figure \ref{simulations1}. It is clear that the major effect of the convolution with a gaussian is to steepen the power spectum on values corresponding to smaller pixel sizes (larger $k$-space values). It is also evident that, even taking into account the reliable region of the power spectrum, there is a certain drift of the power index towards higher values with larger gaussian widths. However, this change is less than 2\% when the beam size changes by a factor of 3, and hence cannot alone explain the large difference between the power index of our observed map and the Kolmogorov regime. Furthermore, as discussed above, errors of this magnitude already arise on different realizations of our simulations.

\begin{figure*}[h]
\begin{center}
\includegraphics[angle=0,scale=0.5,trim=0cm 0cm 0cm 0cm,clip=true]{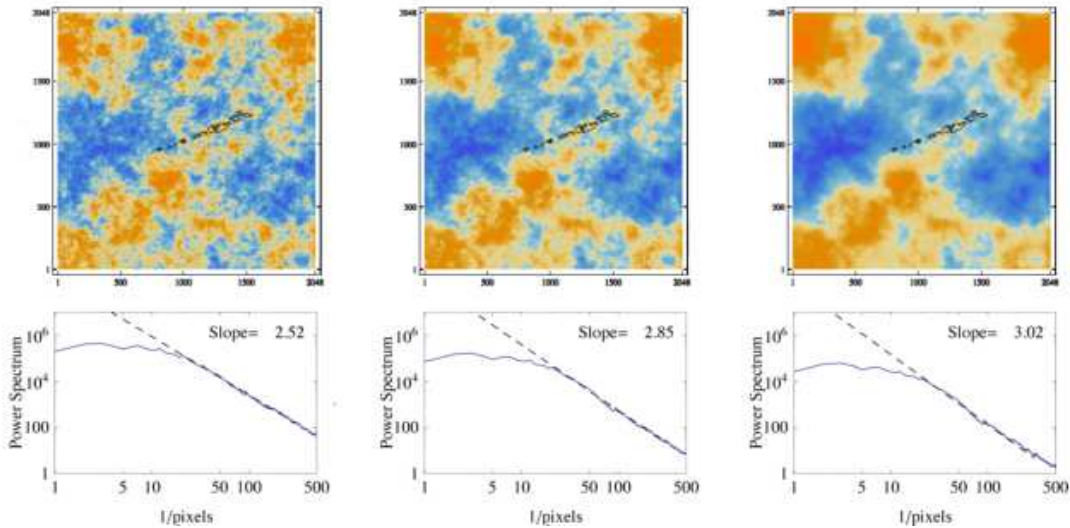}
\caption{Blanking of single realizations of random gaussian field with initial power index $a=$2.5 (left), 3.0 (middle) and 3.5 (right). Top: original realizations, with a black contour superimposed showing the clipping of the data to emulate that of the actual observation. Bottom: power spectrum. Power index fitted to the reliable region of the slope is shown in the top right corner of each graphic. Fit is shown by the dashed line.}
\label{simulations2}
\end{center}
\end{figure*}

In order to investigate how the blanking and data clipping affected the power spectrum of the map, we used some realizations of a simulated gaussian random field maps with power indices 2.5, 3.0 and 3.5. We then blanked the same pixels in these maps as in our original observed RM map. As the blanking process may severely affect the RM distribution, we carefully selected these maps where the resulting blanked RM simulations still followed a gaussian distribution as close as possible. We then produced power spectra from these maps and fitted for the spectral index in the usual way. Resulting maps for the various power indices including a layout of its clipping, and the fitted power spectra are shown in Figure \ref{simulations2}. 

An evident consequence of the blanking of the RM map is the decrease of the power spectrum index values. We note however that the amount of this flattening seems to have a non-linear  dependence on the original power spectrum. Indeed, with an original power spectrum with $a\sim2.5$, the resulting blanked spectrum does not significantly change, whereas for an original power spectrum of  $a\sim3.0$ the flattening is of about 5\% and for the larger power spectrum with $a\sim3.5$, the flattening goes up to more than 15\%. This effect can be explained by inspecting the morphology of the RM realizations for various values of $a$: these realizations with lower $a$ values have much richer morphology at smaller scales (i.e., larger values in the $k$-space) that will remain after clipping. On the other hand, realizations with a larger $a$ value will have their richest features at comparatively larger scales, which are lost when clipped, thus flattening more dramatically the power spectra.

It seems that, in general,  blanking has a comparatively larger effect on the power index compared to convolution, as seen above. Albeit the former can be quite significant, we note that we can still define some limits of applicability. The most straightforward result is that the power index of a map following an original Kolmogorov power spectrum appears to be always larger than $a>3.1$ after clipping. As a consequence, we can not interpret the observed RM as due to a Kolmogorov distribution, and attribute the found comparatively lower power index $a\sim2.5$ to the effects of blanking alone.

\end{document}